\documentclass[runningheads]{llncs}
\usepackage[T1]{fontenc}
\usepackage{graphicx}
\usepackage[caption=false]{subfig}
\usepackage[font=small,labelfont=bf,labelsep=period]{caption}

\begin{document}
\title{AIANO: Enhancing Information Retrieval with AI-Augmented Annotation}

\author{Sameh Khattab\inst{1}\and
Marie Bauer\inst{1}\and
Lukas Heine\inst{1}\and
Till Rostalski\inst{1}\and
Jens Kleesiek\inst{1}\and
Julian Friedrich\inst{1}}
\institute{Institute for Artificial Intelligence in Medicine (IKIM), University Hospital Essen (AöR), Essen, Germany
\newline
\email{\{firstname.lastname\}@uk-essen.de}}
%
%
\authorrunning{S. Khattab et al.}
\maketitle              
\begin{abstract}

The rise of Large Language Models (LLMs) and Retrieval-Augmented Generation (RAG) has rapidly increased the need for high-quality, curated information retrieval datasets. These datasets, however, are currently created with off-the-shelf annotation tools that make the annotation process complex and inefficient. To streamline this process, we developed a specialized annotation tool - AIANO. By adopting an AI-augmented annotation workflow that tightly integrates human expertise with LLM assistance, AIANO enables annotators to leverage AI suggestions while retaining full control over annotation decisions. In a within-subject user study ($n = 15$), participants created question-answering datasets using both a baseline tool and AIANO. AIANO nearly doubled annotation speed compared to the baseline while being easier to use and improving retrieval accuracy. These results demonstrate that AIANO's AI-augmented approach accelerates and enhances dataset creation for information retrieval tasks, advancing annotation capabilities in retrieval-intensive domains.


\end{abstract}

\begin{figure*}[t] 
  \centering
  \includegraphics[
    page=3,
    width=\textwidth,  
    trim=50 100 50 100, 
    clip
  ]{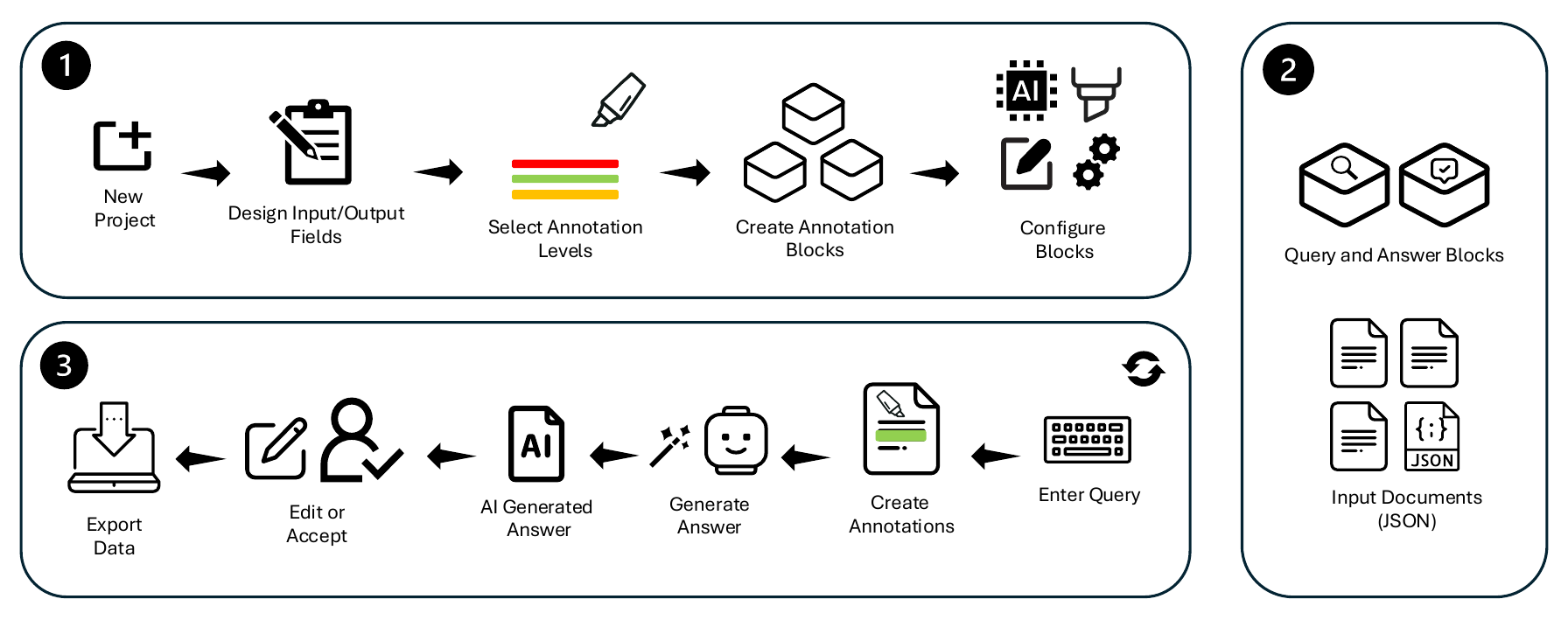}
 \caption{Workflow of the AIANO annotation system. (i) Project Creation Phase: Configure project metadata, input/output schemas, annotation levels, and AIANO Blocks. (ii) Project Configuration Phase: Configure annotation blocks with LLM provider and upload documents for annotation. (iii) Annotation Phase: Annotators highlight text, trigger AI-assisted content generation, review, edit, and export the dataset. The cycle icon indicates iterative refinement.}
  \label{fig:aiano-overview}
\end{figure*}

\section{Introduction}

Large Language Models (LLMs) have become integral to workflows across diverse domains \cite{attentionisalluneed}, yet remain prone to reliability issues and factual inaccuracies \cite{hallucination}. These limitations can be mitigated by providing relevant documents as context, a technique called Retrieval-Augmented Generation (RAG) \cite{rag}. However, the effectiveness of RAG systems depends on multiple components \cite{izacard2021leveragingpassageretrievalgenerative}, making robust evaluation with high-quality annotated datasets essential \cite{ragbench,es2025ragasautomatedevaluationretrieval}. Yet, these datasets require queries, ground-truth answers, and relevant document annotations, making their creation - particularly in specialized domains - prohibitively challenging, time-consuming, and expensive  \cite{ares,sorodoc2025garagebenchmarkgroundingannotations,wang2025omnievalomnidirectionalautomaticrag}.  

To streamline the creation of these datasets, we developed AIANO \footnote{ AIANO is available at https://github.com/TIO-IKIM/AIANO}, to our knowledge the first annotation tool designed specifically for information retrieval (IR) tasks.  AIANO implements an AI-augmented workflow that seamlessly integrates human expertise with LLM capabilities, accelerating annotation through AI assistance while maintaining quality through human oversight. We make three primary contributions:
\begin{itemize}
    \item We introduce AIANO, a specialized annotation tool for information retrieval tasks.
    \item We outline AIANO's design principles enabling efficient LLM-human collaboration in semi-automatic annotation workflows.
    \item We systematically evaluate AIANO's effectiveness and usability against a baseline annotation tool through a controlled within-subject user study.
\end{itemize} 

\section{Related Work}   

Creating IR datasets presents unique challenges that distinguish it from standard tasks such as classification or entity recognition. IR annotation requires synthesizing information across multiple documents and constructing contextual rationales, making the process time-intensive and prone to incomplete coverage \cite{informationretrievallacksinform}. This incompleteness carries significant consequences: research demonstrates that when annotators miss relevant passages, retrieval models trained on these incomplete annotations suffer substantial performance degradation \cite{rassin2024evaluatingdmeritpartialannotationinformation}.  

The emergence of RAG systems has further intensified IR annotation challenges, as RAG relies on retrieval mechanisms to ground LLM responses in external knowledge \cite{rag}. This reliance has spurred the development of evaluation metrics for assessing RAG system performance across multiple dimensions \cite{wang2025omnievalomnidirectionalautomaticrag,ragbench,es2025ragasautomatedevaluationretrieval}. However, RAG evaluation frameworks require high-quality annotated datasets, which remain expensive and labor-intensive to produce.

The high cost of annotation has motivated research into leveraging LLMs themselves for automatic data labeling. Such automation efforts have shown promise, with studies demonstrating that models like GPT-3.5 can achieve human-comparable performance on certain tasks using explain-then-annotate approaches \cite{he2024annollmmakinglargelanguage}. However, this performance varies substantially: research across several annotation tasks with different datasets reveals significant dependence on task type and domain, necessitating human validation to ensure data quality and trustworthiness \cite{pangakis2023automatedannotationgenerativeai}. This variability in automated annotation quality has motivated human-AI collaborative frameworks that leverage complementary strengths of both parties \cite{AI-Assistedlabeling,patat,li-etal-2023-coannotating}. For instance, CoAnnotating \cite{li-etal-2023-coannotating}, employs uncertainty estimation to strategically allocate annotation instances, routing low-uncertainty cases to LLMs while directing uncertain instances to humans; thereby achieving performance improvements over purely manual or fully automated methods. However, these automation-focused solutions do not fully address the fundamental workflow inefficiencies inherent in document-intensive IR annotation, such as managing multi-document contexts, ensuring comprehensive coverage, and balancing annotation speed with quality. 

\noindent AIANO tackles these challenges through a task-specialized, semi-automatic pipeline that strategically integrates LLM assistance while preserving human oversight in labeling decisions.

\section{AIANO System} 

AIANO (\textbf{AI A}ugmented an\textbf{NO}tation) provides a platform for dataset creation through human-AI collaboration, designed primarily for information retrieval tasks but adaptable to diverse annotation scenarios. 

\subsection{Core Concepts} 

\subsubsection{AIANO Blocks}  AIANO models annotation tasks as configurable input/output blocks, each operating in one of three modes that represent varying levels of human-AI collaboration:

\textbf{(i) Plain Mode} receives no automatic input sources. The AI performs no operations, and the human annotator manually writes all content from scratch. For example, a free-text Comment Block would allow annotators to write notes directly. 

\textbf{(ii) AI Solo Mode} takes pre-defined system prompts as input. The AI automatically generates content based on these prompts, which the human annotator can then review and refine. For example, a Question Block might auto-generate boilerplate comprehension questions. 

\textbf{(iii) Human-AI Collaborative Mode} draws from multiple input sources: existing annotations, user-defined fields, other blocks, and system prompts. The AI generates outputs by synthesizing these sources, and the human annotator can accept, modify, or override the suggestions. For example, an Answer Block may draw from a Question Block, highlighted passages, and document metadata to generate candidate answers. Users can also create custom block types for specialized needs. 

\subsubsection{Annotation Levels} AIANO supports configurable annotation levels for highlighting text with different categories (e.g. "important", "distracting"), providing contextual information for downstream tasks.  

\subsubsection{Input Schema Flexibility} 

 Users define custom input and output schemas through the UI, following a JSON structure that requires only document ID and subject ID as mandatory fields. Additional fields of any type can be added, enabling support for varying document types without requiring programmatic configuration. 

\subsection{Workflow} 

To illustrate AIANO's capabilities, creating a question-answer dataset for RAG evaluation shall be considered. The workflow comprises three phases: \textbf{(1) Project Creation}: configure metadata, define schemas, set annotation levels, and design tasks using AIANO Blocks; \textbf{(2) Configuration}: connect blocks to LLMs and upload JSON documents; \textbf{(3) Annotation}: iteratively highlight text, generate an answer, review content, and export datasets with full provenance. The detailed workflow is shown in Fig.\ref{fig:aiano-overview}.

The project setup follows four steps: define metadata (name, description, tags), configure input/output schemas and upload documents, set annotation levels (e.g., highlight levels for evidence passages), and configure AIANO Blocks (type, mode, inputs, prompts). For RAG datasets, a Question Block in Plain Mode enables manual formulation, while an Answer Block in Human-AI Collaborative Mode generates answers from questions and highlights. 

The annotation interface comprises three panels:  document corpus with search and filtering capabilities (left), highlighting interface with search and annotation tools (center), and AIANO Blocks (right). Annotators select documents, highlight relevant spans, and populate blocks left-to-right, and the system automatically saves annotations with provenance metadata for the current entry as well as the previous annotation entries for the project. 

For downstream applications, users can export datasets in JSON format with question-answer-passage triplets, IDs, and span positions. Projects export in \textit{.aiano} format encapsulating all configurations for reproducibility and sharing. 

\subsection{Implementation} 

AIANO supports any LLM provider following OpenAI API \cite{openaiapi} standards, including commercial services (e.g., OpenAI, Anthropic) and local deployments such as vLLM \cite{kwon2023efficient} for efficient inference, enabling cost-effective, high-throughput workflows. The system uses a containerized microservices architecture with React 19 frontend, FastAPI backend, PostgreSQL database, and Docker deployment.

\section{Methodology} 

\subsection{User Study} 

\subsubsection{Study Design}  

To evaluate AIANO's effectiveness over existing annotation tools, we conducted a within-subject study comparing AIANO to Label Studio \cite{labelstudio}, a widely used annotation tool. We hypothesized that AIANO would demonstrate lower cognitive load, faster task completion, and higher retrieval accuracy. 

\subsubsection{Participants} We recruited 15 participants, including graduate students, researchers, software developers, medical doctors, and regulatory affairs specialists (median age 26, range 18-50; 66.7\% men, 33.3\% women; all German speakers with varying annotation experience). Participants received no compensation. 

\subsubsection{Experimental Setup} We created 60 short German general knowledge documents (3-5 sentences each) using Claude Sonnet 4, then authored nine questions requiring either single-document (n=5) or multi-document retrieval (n=4, requiring 3-4 documents). Participants received predefined questions and searched for answers by identifying relevant documents, highlighting passages, and formulating responses. Each participant completed four questions per platform (two single-document, two multi-document, randomly selected) without time limits. 

To mitigate order and carryover effects, participants were randomly assigned to start with either tool, given approximately 10-minute breaks between platforms, and approximately half received identical questions across tools, while others received different questions of comparable difficulty. 

\textbf{AIANO} was configured with a Question Block (Plain Mode) and Answer Block (Human-AI Collaborative Mode), where participants accessed Meta's Llama 70B model \cite{touvron2023llamaopenefficientfoundation} to generate answers from highlighted passages. AIANO provided full-text search across and within documents. \textbf{Label Studio} (v1.13.1) was configured with separate projects per question but lacked native interactive AI collaboration and full-text search capabilities. 

We conducted a brief tutorial that included written instructions, demonstrations, and hands-on practice before the task began. We measured task completion time, collected NASA-TLX assessments \cite{HART1988139} and post-task questionnaires for each tool, and recorded the annotated datasets.  

\subsection{Evaluation Metrics} 

\subsubsection{Subjective Measures} Participants completed questionnaires after each platform with eight questions assessing usability, navigation, and performance (rated on 5-point Likert scales), from which we calculated a \textit{Composite} usability score. For AIANO, we included an additional question assessing AI-assisted features and search functionality. In addition, we measured cognitive load using NASA-TLX dimensions: \textit{Temporal Demand}, \textit{Physical Demand}, \textit{Mental Demand}, \textit{Frustration}, \textit{Performance}, and \textit{Effort}. Open-ended questions gathered qualitative feedback. 

\subsubsection{Objective Measures} We measured task completion time (from start to participant-indicated completion) and calculated retrieval metrics (precision, recall, and F1 score) by comparing retrieved documents (those with highlights) against predefined relevant documents. A document was considered retrieved if any passage within it was highlighted. 

\subsubsection{Statistical Analysis} We assessed normality using Shapiro-Wilk tests \cite{shapiro}. Most subjective measures and task completion time followed normal distributions, while \textit{Mental Demand}, \textit{Physical Demand}, \textit{Temporal Demand}, \textit{Reuse Intention}, \textit{Speed}, and \textit{Overall Satisfaction} violated normality. We applied paired t-tests for normally distributed data and Wilcoxon signed-rank tests for non-parametric data ($p < 0.05$), with $p < 0.001$ for values below this threshold. Statistics are reported as medians unless indicated otherwise.

\begin{figure}[htbp]
  \centering
  \begin{minipage}[c]{0.48\textwidth}
    \centering
    \includegraphics[page=1, width=\textwidth]{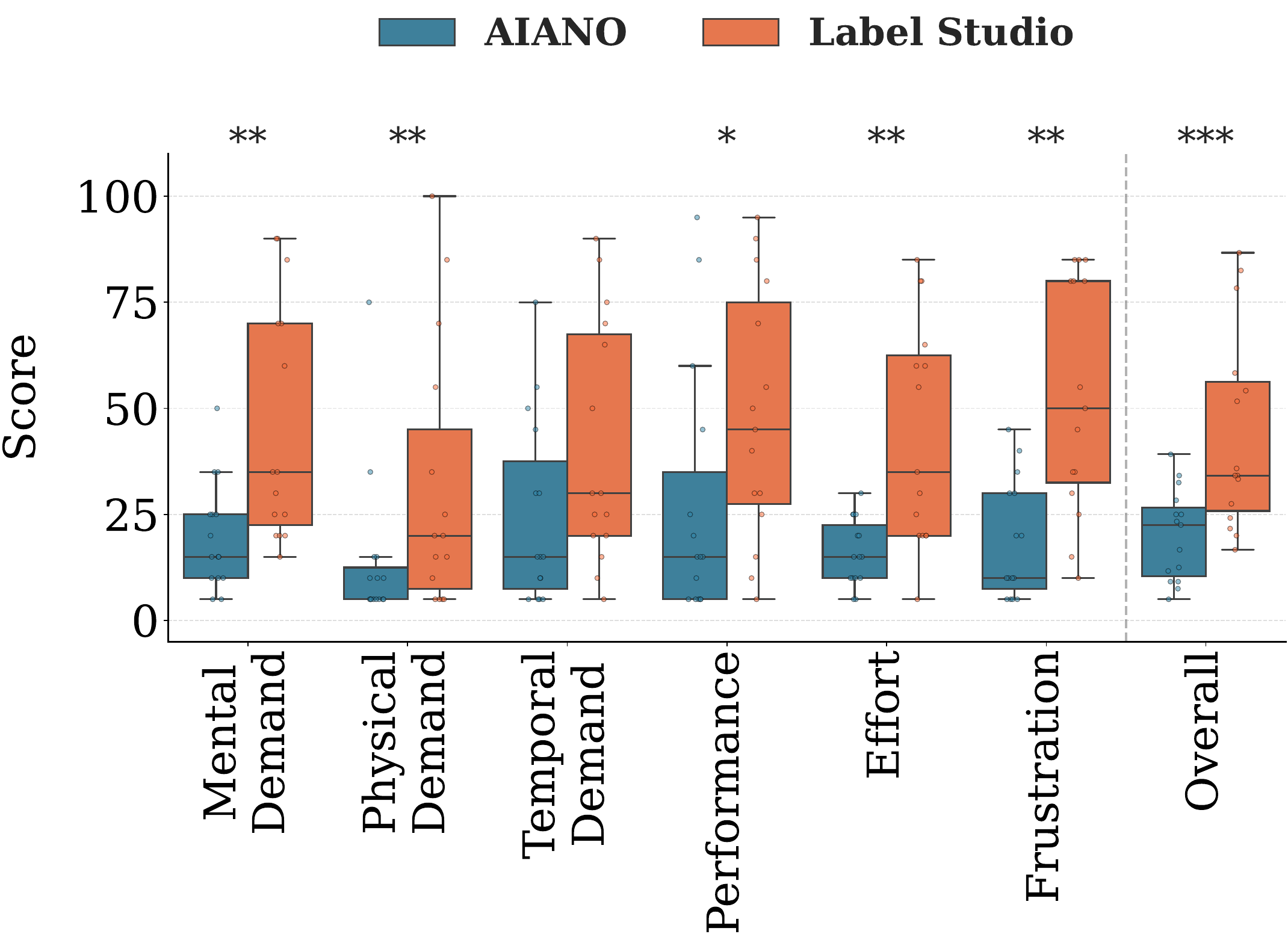}
    \caption{NASA-TLX workload assessment. \textmd{Subscale scores across six dimensions and overall workload. Lower scores indicate lower workload. $* p < 0.05, ** p < 0.01, *** p < 0.001$.}}
    \label{fig:nasa}
  \end{minipage}
  \hfill
  \begin{minipage}[c]{0.48\textwidth}
    \centering
    \includegraphics[page=1, width=\textwidth]{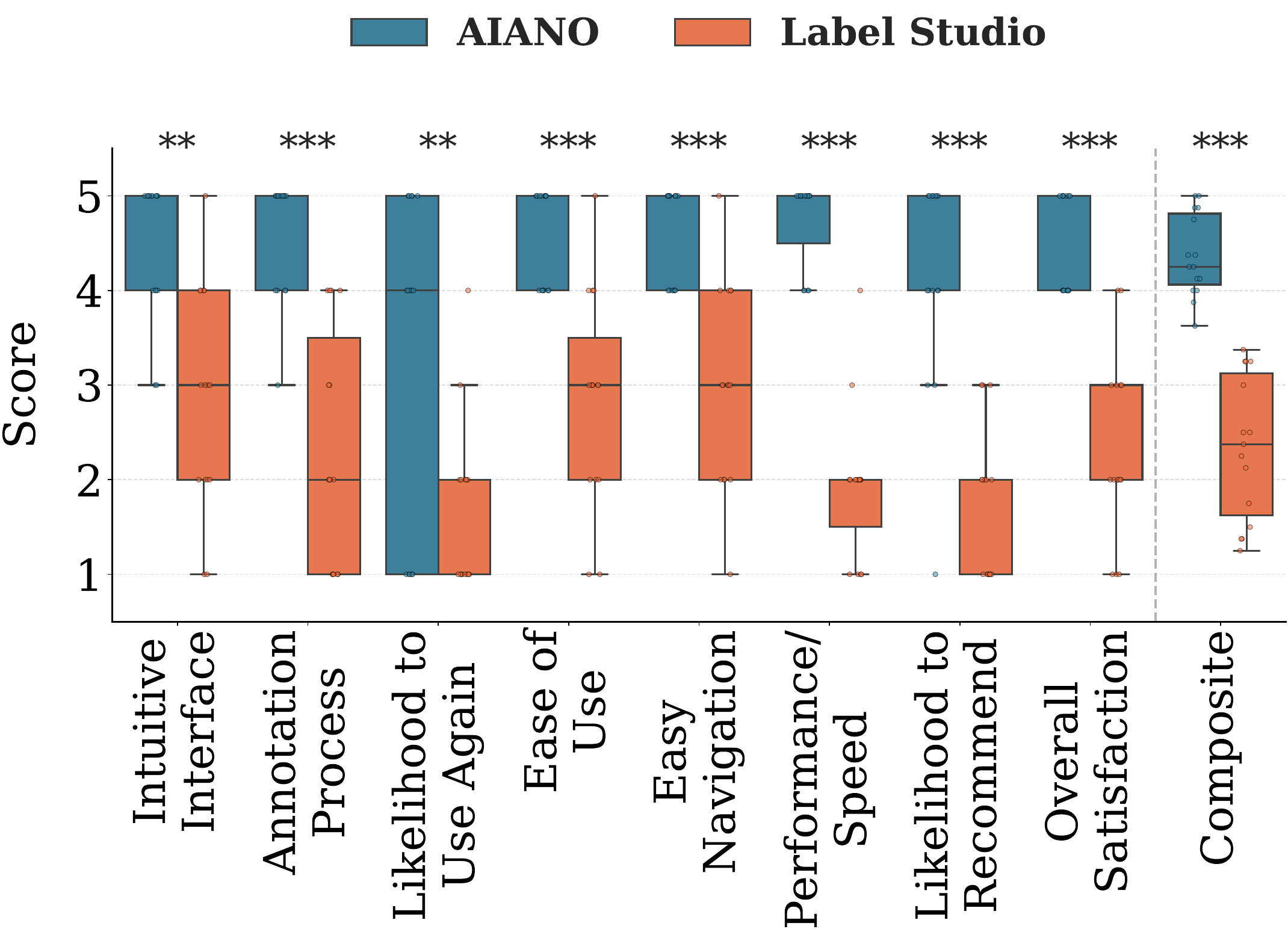}
    \caption{Usability questionnaire ratings. \textmd{Likert scale ratings ($1-5$) across eight usability dimensions and a composite score. Higher scores indicate better user experience. $** p < 0.01, *** p < 0.001$.}}
    \label{fig:quiestionnaire}
  \end{minipage}
\end{figure}

\begin{figure}[htbp]
  \centering
  \begin{minipage}[c]{0.48\textwidth}
    \centering
    \includegraphics[page=1, width=\textwidth]{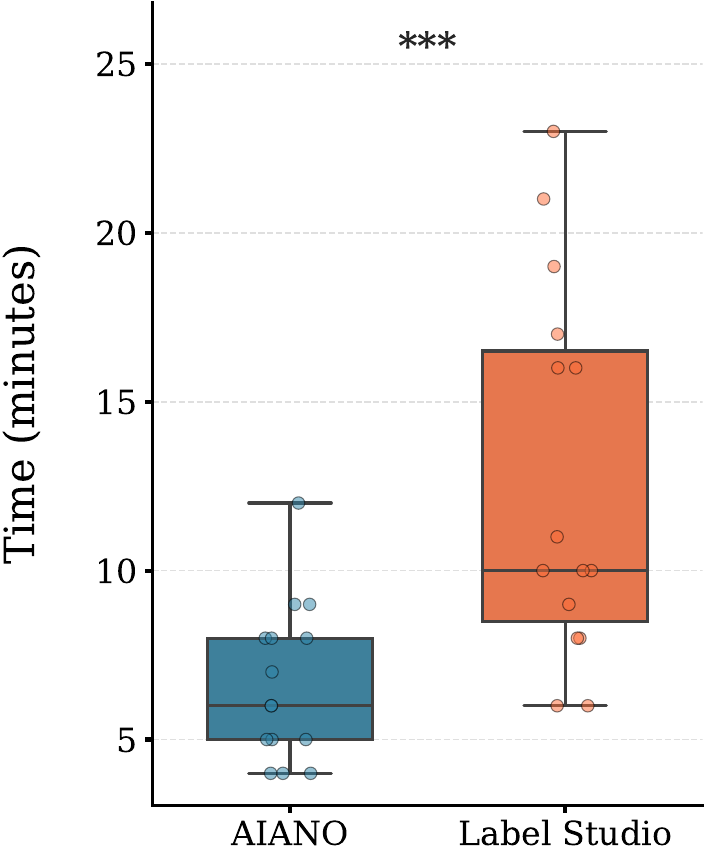}
    \caption{Task completion time. \textmd{Individual participant task durations (in minutes) are shown as overlaid circles ($n = 15$).}}
    \label{fig:completion_time}
  \end{minipage}
  \hfill
  \begin{minipage}[c]{0.48\textwidth}
    \centering
    \captionof{table}{Information retrieval performance comparison.}
    \label{tab:retrieval_comparison}
    \begin{tabular}{|l|c|c|}
      \hline
      Metric & AIANO & Label Studio \\
      \hline
      Precision & 0.889 & 0.867 \\
      Recall & 0.883 & 0.783 \\
      F1-Score & 0.860 & 0.787 \\
      \hline
    \end{tabular}
  \end{minipage}
\end{figure}

\section{Results and Analysis}  

\subsection{Quantitative Analysis}  

\subsubsection{Post-Study Questionnaires}  

Participants reported lower overall workload with AIANO compared to Label Studio (22.5 vs. 34.17, p < 0.001). Across NASA-TLX dimensions (see Fig.\ref{fig:nasa}), AIANO consistently reduced mental demand (15.0 vs. 35.0, p = 0.005), physical demand (5.0 vs. 20.0, p = 0.008), required less effort (15.0 vs. 35.0, p = 0.003), and lowered frustration (10.0 vs. 50.0, p = 0.001). Participants also rated their performance better with AIANO (15.0 vs. 45.0, p = 0.037). Temporal demand did not differ between tools (15.0 vs. 30.0, p = 0.083).  

Feedback from the custom 5-point Likert questionnaire paralleled NASA-TLX results. All participants rated AIANO with scores of 4 or 5, while Label Studio exhibited greater variability and lower ratings. Participants consistently rated AIANO higher across all usability dimensions, including intuitive interface (5.0 vs. 3.0, p = 0.001), annotation process (5.0 vs. 2.0, p < 0.001), likelihood to use again (4.0 vs. 1.0, p = 0.008), ease of use (5.0 vs. 3.0, p < 0.001), easy navigation (5.0 vs. 3.0, p < 0.001), performance/speed (5.0 vs. 2.0, p < 0.001), likelihood to recommend (4.0 vs. 2.0, p < 0.001), and overall satisfaction (4.0 vs. 2.0, p < 0.001). The composite score demonstrated higher usability for AIANO (4.25 vs. 2.375, p < 0.001). Most participants rated AI-assisted features highly (5.0), with no Label Studio comparison possible due to lack of this functionality.  

\subsubsection{Task Completion Time}  

Participants completed annotation tasks in nearly half the time with AIANO compared to Label Studio (Fig.\ref{fig:completion_time}). Median task completion time decreased from 10.0 minutes with Label Studio to 6.0 minutes with AIANO.  

\subsubsection{Retrieval Performance}  

AIANO demonstrated higher retrieval performance across all metrics (see Table \ref{tab:retrieval_comparison}). Participants using AIANO achieved higher precision (0.89 vs. 0.87), recall (0.88 vs. 0.78), and F1 score (0.86 vs. 0.79), with an average improvement of 8.2\% across the three metrics (precision: +2.5\%, recall: +12.8\%, F1: +9.3\%).  

\subsection{Qualitative Feedback} Participants found AIANO intuitive and efficient, reporting that AI-assisted answer generation and search functionality simplified workflows and accelerated task completion. Specifically, 86.7\% found search useful, 93.3\% found AI assistance useful, and 80\% valued both features for improved efficiency.  

In contrast, Label Studio feedback highlighted performance issues when opening/closing documents and difficulties copying text or navigating between documents. Some participants found Label Studio adequate, appreciating text preview and the organized question-answer layout.  

\section{Discussion}In this work, we introduced AIANO, an AI-augmented annotation tool, and demonstrated through a comparative study with Label Studio that AIANO substantially accelerates the creation of IR datasets while reducing cognitive load and improving usability. These results show that dataset creation of IR tasks can be significantly enhanced by strategically integrating annotation workflows with AI assistance. 

\subsection{Efficiency and Usability} In the user study, participants reported substantially lower workload with AIANO, with significant reductions across mental demand, physical demand, effort, and frustration. The reduction in frustration is particularly noteworthy, as high frustration contributes to burnout, reduced data quality, and higher turnover in annotation work. Participants completed tasks 40\% faster without reporting increased time pressure, indicating genuine efficiency gains rather than rushing. These improvements likely stem from integrated full-text search and AI-assisted answer generation, which transformed document discovery and reduced formulation effort. Participants consistently rated AIANO highly on interface intuitiveness, ease of use, and navigation. 

\subsection{Retrieval Performance and Data Quality} 

Annotators using AIANO identified more of the truly relevant documents in the corpus, as indicated by higher recall, while simultaneously reducing false positives, reflected in improved precision. The substantial gains in F1 scores confirm AIANO's overall effectiveness in balancing comprehensiveness with accuracy. This directly tackles a critical bottleneck: incomplete and noisy annotations lead to biased training data and unreliable model evaluation \cite{rassin2024evaluatingdmeritpartialannotationinformation}. AIANO's improved coverage and discrimination enable the creation of datasets that better capture the full range of relevant information while filtering out noise, ultimately producing more robust retrieval systems. 

\subsection{LLM-Assisted Annotation} 

The integrated LLM assistance and search functionality worked synergistically throughout the annotation workflow: participants searched for documents, highlighted passages, and used AI to generate candidate answers. This represents a fundamentally different annotation paradigm compared to manual approaches. Qualitative feedback confirmed the value: 86.7\% found search useful, 93.3\% found AI assistance useful, and 80\% valued both together for improved efficiency. This aligns with perspectives on human-AI collaborative annotation \cite{humanaicollaboration}, positioning LLMs as interactive collaborators embedded directly in the annotation workflow rather than preprocessing tools applied before human involvement. However, as \cite{pangakis2023automatedannotationgenerativeai} noted, LLM effectiveness can be task-dependent, warranting investigation across diverse annotation scenarios to understand when and how LLM assistance provides the most value. 

\subsection{Limitations} 

 Several limitations should be acknowledged. Our user study included 15 participants, which constrains the statistical power for subgroup analyses and limits our ability to detect nuanced effects across user types. Additionally, comparing AIANO against a single baseline prevents us from situating its performance relative to the broader landscape of annotation platforms. Our evaluation centered on German-language IR annotation with short documents; broader validation is needed across different annotation types (named entity recognition, sentiment analysis), varying document lengths, diverse domains (medical, legal), and multiple languages to establish generalizability. Finally, we did not capture detailed usage patterns for AI assistance or search functionality, assess answer quality beyond retrieval metrics, or systematically investigate how task characteristics, such as single versus multi-document questions influence annotation outcomes. 

\subsection{Future Directions} 

Although AIANO was designed specifically for IR dataset creation, its approach of tightly integrating large language models into the annotation workflow where LLMs actively assist in generating, suggesting, and refining annotations across flexible block-based structures suggests broader applicability to annotation contexts requiring multi-document navigation and information synthesis. This LLM-augmented approach positions AIANO as a middle ground between traditional annotation tools that lack intelligent assistance and overly generic platforms without domain-specific affordances. Future research should investigate how LLM assistance scales across diverse annotation tasks and domains, examine potential drawbacks such as automation bias or annotator deskilling, and develop evidence-based guidelines for effective human-AI collaboration in dataset creation workflows.

\section{Conclusion} 

This paper introduces AIANO, a specialized annotation tool that natively integrates AI assistance and full-text search into the information retrieval annotation workflow. Our evaluation shows that AIANO reduces cognitive workload, accelerates task completion, and improves retrieval performance compared to general-purpose annotation tools. By addressing the specific demands of retrieval-intensive annotation tasks, AIANO enables more efficient and effective creation of information retrieval datasets, thus advancing annotation capabilities in retrieval-intensive domains.

\begin{credits}
\subsubsection{\ackname}
The authors acknowledge using Cursor for code assistance, ChatGPT-4 for demo dataset generation, and Claude Sonnet 4 for text paraphrasing. All AI-generated outputs were reviewed, verified, and edited by the authors to ensure accuracy and originality. 
We sincerely thank all participants for taking part in our user study and for their constructive feedback. 

\subsubsection{\discintname}
The authors have no competing interests to declare that are relevant to the content of this article.
\end{credits}
%
\bibliographystyle{splncs04}
\bibliography{aiano}

@misc{he2024annollmmakinglargelanguage,
      title={AnnoLLM: Making Large Language Models to Be Better Crowdsourced Annotators}, 
      author={Xingwei He and Zhenghao Lin and Yeyun Gong and A-Long Jin and Hang Zhang and Chen Lin and Jian Jiao and Siu Ming Yiu and Nan Duan and Weizhu Chen},
      year={2024},
      eprint={2303.16854},
      archivePrefix={arXiv},
      primaryClass={cs.CL},
      url={https://arxiv.org/abs/2303.16854}, 
}

@misc{pangakis2023automatedannotationgenerativeai,
      title={Automated Annotation with Generative AI Requires Validation}, 
      author={Nicholas Pangakis and Samuel Wolken and Neil Fasching},
      year={2023},
      eprint={2306.00176},
      archivePrefix={arXiv},
      primaryClass={cs.CL},
      url={https://arxiv.org/abs/2306.00176}, 
}

@misc{rassin2024evaluatingdmeritpartialannotationinformation,
      title={Evaluating D-MERIT of Partial-annotation on Information Retrieval}, 
      author={Royi Rassin and Yaron Fairstein and Oren Kalinsky and Guy Kushilevitz and Nachshon Cohen and Alexander Libov and Yoav Goldberg},
      year={2024},
      eprint={2406.16048},
      archivePrefix={arXiv},
      primaryClass={cs.IR},
      url={https://arxiv.org/abs/2406.16048}, 
}

@inproceedings{patat,
author = {Gebreegziabher, Simret Araya and Zhang, Zheng and Tang, Xiaohang and Meng, Yihao and Glassman, Elena L. and Li, Toby Jia-Jun},
title = {PaTAT: Human-AI Collaborative Qualitative Coding with Explainable Interactive Rule Synthesis},
year = {2023},
isbn = {9781450394215},
publisher = {Association for Computing Machinery},
address = {New York, NY, USA},
url = {https://doi.org/10.1145/3544548.3581352},
doi = {10.1145/3544548.3581352},
booktitle = {Proceedings of the 2023 CHI Conference on Human Factors in Computing Systems},
articleno = {362},
numpages = {19},
location = {Hamburg, Germany},
series = {CHI '23}
}

@inproceedings{li-etal-2023-coannotating,
    title = "{C}o{A}nnotating: Uncertainty-Guided Work Allocation between Human and Large Language Models for Data Annotation",
    author = "Li, Minzhi  and
      Shi, Taiwei  and
      Ziems, Caleb  and
      Kan, Min-Yen  and
      Chen, Nancy  and
      Liu, Zhengyuan  and
      Yang, Diyi",
    editor = "Bouamor, Houda  and
      Pino, Juan  and
      Bali, Kalika",
    booktitle = "Proceedings of the 2023 Conference on Empirical Methods in Natural Language Processing",
    month = dec,
    year = "2023",
    address = "Singapore",
    publisher = "Association for Computational Linguistics",
    url = "https://aclanthology.org/2023.emnlp-main.92/",
    doi = "10.18653/v1/2023.emnlp-main.92",
    pages = "1487--1505"
}

@misc{rag,
      title={Retrieval-Augmented Generation for Knowledge-Intensive NLP Tasks}, 
      author={Patrick Lewis and Ethan Perez and Aleksandra Piktus and Fabio Petroni and Vladimir Karpukhin and Naman Goyal and Heinrich Küttler and Mike Lewis and Wen-tau Yih and Tim Rocktäschel and Sebastian Riedel and Douwe Kiela},
      year={2021},
      eprint={2005.11401},
      archivePrefix={arXiv},
      primaryClass={cs.CL},
      url={https://arxiv.org/abs/2005.11401}, 
}

@misc{attentionisalluneed,
      title={Attention Is All You Need}, 
      author={Ashish Vaswani and Noam Shazeer and Niki Parmar and Jakob Uszkoreit and Llion Jones and Aidan N. Gomez and Lukasz Kaiser and Illia Polosukhin},
      year={2023},
      eprint={1706.03762},
      archivePrefix={arXiv},
      primaryClass={cs.CL},
      url={https://arxiv.org/abs/1706.03762}, 
}

@inproceedings{humanaicollaboration,
author = {Wang, Dakuo and Churchill, Elizabeth and Maes, Pattie and Fan, Xiangmin and Shneiderman, Ben and Shi, Yuanchun and Wang, Qianying},
title = {From Human-Human Collaboration to Human-AI Collaboration: Designing AI Systems That Can Work Together with People},
year = {2020},
isbn = {9781450368193},
publisher = {Association for Computing Machinery},
address = {New York, NY, USA},
url = {https://doi.org/10.1145/3334480.3381069},
doi = {10.1145/3334480.3381069},
booktitle = {Extended Abstracts of the 2020 CHI Conference on Human Factors in Computing Systems},
pages = {1–6},
numpages = {6},
location = {Honolulu, HI, USA},
series = {CHI EA '20}
}

@article{AI-Assistedlabeling,
author = {Ashktorab, Zahra and Desmond, Michael and Andres, Josh and Muller, Michael and Joshi, Narendra Nath and Brachman, Michelle and Sharma, Aabhas and Brimijoin, Kristina and Pan, Qian and Wolf, Christine T. and Duesterwald, Evelyn and Dugan, Casey and Geyer, Werner and Reimer, Darrell},
title = {AI-Assisted Human Labeling: Batching for Efficiency without Overreliance},
year = {2021},
issue_date = {April 2021},
publisher = {Association for Computing Machinery},
address = {New York, NY, USA},
volume = {5},
number = {CSCW1},
url = {https://doi.org/10.1145/3449163},
doi = {10.1145/3449163},
journal = {Proc. ACM Hum.-Comput. Interact.},
month = apr,
articleno = {89},
numpages = {27}
}

@misc{es2025ragasautomatedevaluationretrieval,
      title={Ragas: Automated Evaluation of Retrieval Augmented Generation}, 
      author={Shahul Es and Jithin James and Luis Espinosa-Anke and Steven Schockaert},
      year={2025},
      eprint={2309.15217},
      archivePrefix={arXiv},
      primaryClass={cs.CL},
      url={https://arxiv.org/abs/2309.15217}, 
}

@misc{ragbench,
      title={RAGBench: Explainable Benchmark for Retrieval-Augmented Generation Systems}, 
      author={Robert Friel and Masha Belyi and Atindriyo Sanyal},
      year={2025},
      eprint={2407.11005},
      archivePrefix={arXiv},
      primaryClass={cs.CL},
      url={https://arxiv.org/abs/2407.11005}, 
}

@misc{wang2025omnievalomnidirectionalautomaticrag,
      title={OmniEval: An Omnidirectional and Automatic RAG Evaluation Benchmark in Financial Domain}, 
      author={Shuting Wang and Jiejun Tan and Zhicheng Dou and Ji-Rong Wen},
      year={2025},
      eprint={2412.13018},
      archivePrefix={arXiv},
      primaryClass={cs.CL},
      url={https://arxiv.org/abs/2412.13018}, 
}

@misc{ares,
      title={ARES: An Automated Evaluation Framework for Retrieval-Augmented Generation Systems}, 
      author={Jon Saad-Falcon and Omar Khattab and Christopher Potts and Matei Zaharia},
      year={2024},
      eprint={2311.09476},
      archivePrefix={arXiv},
      primaryClass={cs.CL},
      url={https://arxiv.org/abs/2311.09476}, 
}

@misc{sorodoc2025garagebenchmarkgroundingannotations,
      title={GaRAGe: A Benchmark with Grounding Annotations for RAG Evaluation}, 
      author={Ionut-Teodor Sorodoc and Leonardo F. R. Ribeiro and Rexhina Blloshmi and Christopher Davis and Adrià de Gispert},
      year={2025},
      eprint={2506.07671},
      archivePrefix={arXiv},
      primaryClass={cs.CL},
      url={https://arxiv.org/abs/2506.07671}, 
}

@incollection{HART1988139,
title = {Development of NASA-TLX (Task Load Index): Results of Empirical and Theoretical Research},
editor = {Peter A. Hancock and Najmedin Meshkati},
series = {Advances in Psychology},
publisher = {North-Holland},
volume = {52},
pages = {139-183},
year = {1988},
booktitle = {Human Mental Workload},
issn = {0166-4115},
doi = {https://doi.org/10.1016/S0166-4115(08)62386-9},
url = {https://www.sciencedirect.com/science/article/pii/S0166411508623869},
author = {Sandra G. Hart and Lowell E. Staveland}
}

@article{hallucination,
   title={Survey of Hallucination in Natural Language Generation},
   volume={55},
   ISSN={1557-7341},
   url={http://dx.doi.org/10.1145/3571730},
   DOI={10.1145/3571730},
   number={12},
   journal={ACM Computing Surveys},
   publisher={Association for Computing Machinery (ACM)},
   author={Ji, Ziwei and Lee, Nayeon and Frieske, Rita and Yu, Tiezheng and Su, Dan and Xu, Yan and Ishii, Etsuko and Bang, Ye Jin and Madotto, Andrea and Fung, Pascale},
   year={2023},
   month=mar, pages={1–38} }

@misc{izacard2021leveragingpassageretrievalgenerative,
      title={Leveraging Passage Retrieval with Generative Models for Open Domain Question Answering}, 
      author={Gautier Izacard and Edouard Grave},
      year={2021},
      eprint={2007.01282},
      archivePrefix={arXiv},
      primaryClass={cs.CL},
      url={https://arxiv.org/abs/2007.01282}, 
}

@inproceedings{informationretrievallacksinform,
author = {B\"{u}ttcher, Stefan and Clarke, Charles L. A. and Yeung, Peter C. K. and Soboroff, Ian},
title = {Reliable information retrieval evaluation with incomplete and biased judgements},
year = {2007},
isbn = {9781595935977},
publisher = {Association for Computing Machinery},
address = {New York, NY, USA},
url = {https://doi.org/10.1145/1277741.1277755},
doi = {10.1145/1277741.1277755},
booktitle = {Proceedings of the 30th Annual International ACM SIGIR Conference on Research and Development in Information Retrieval},
pages = {63–70},
numpages = {8},
location = {Amsterdam, The Netherlands},
series = {SIGIR '07}
}

@inproceedings{kwon2023efficient,
  title={Efficient Memory Management for Large Language Model Serving with PagedAttention},
  author={Woosuk Kwon and Zhuohan Li and Siyuan Zhuang and Ying Sheng and Lianmin Zheng and Cody Hao Yu and Joseph E. Gonzalez and Hao Zhang and Ion Stoica},
  booktitle={Proceedings of the ACM SIGOPS 29th Symposium on Operating Systems Principles},
  year={2023}
}

@misc{openaiapi,
  title = {OpenAI API Documentation},
  howpublished = {\url{https://platform.openai.com/docs/api-reference}},
  note = {Accessed: 2025-11-04}
}

@misc{labelstudio,
  title        = {Label Studio},
  howpublished = {\url{https://labelstud.io}},
  note         = {Accessed: 2025-11-04}
}

@misc{touvron2023llamaopenefficientfoundation,
      title={LLaMA: Open and Efficient Foundation Language Models}, 
      author={Hugo Touvron and Thibaut Lavril and Gautier Izacard and Xavier Martinet and Marie-Anne Lachaux and Timothée Lacroix and Baptiste Rozière and Naman Goyal and Eric Hambro and Faisal Azhar and Aurelien Rodriguez and Armand Joulin and Edouard Grave and Guillaume Lample},
      year={2023},
      eprint={2302.13971},
      archivePrefix={arXiv},
      primaryClass={cs.CL},
      url={https://arxiv.org/abs/2302.13971}, 
}

@article{shapiro,
 ISSN = {00063444, 14643510},
 URL = {http://www.jstor.org/stable/2333709},
 author = {S. S. Shapiro and M. B. Wilk},
 journal = {Biometrika},
 number = {3/4},
 pages = {591--611},
 publisher = {[Oxford University Press, Biometrika Trust]},
 title = {An Analysis of Variance Test for Normality (Complete Samples)},
 urldate = {2025-11-05},
 volume = {52},
 year = {1965}
}





\end{document}